\documentclass[ ]{IEEEtran}
\usepackage{amssymb}
\usepackage{amsmath,amsfonts}
\usepackage[ruled,linesnumbered]{algorithm2e}
\usepackage{amsthm}
\usepackage{array}
\usepackage{textcomp}
\usepackage{stfloats}
\usepackage{url}
\usepackage{verbatim}
\usepackage{graphicx}
\usepackage{cite}
\usepackage{bm}
\usepackage{subfigure}
\usepackage{float}
\usepackage{booktabs}
\usepackage{tikz}
\newcommand{\RNum}[1]{\uppercase\expandafter{\romannumeral #1\relax}}

\theoremstyle{definition}

\theoremstyle{remark}

\begin{document}

\title{Generating CKM Using Others' Data: Cross-AP CKM Inference 
with Deep Learning 
}

\author{{Zhuoyin Dai,  Di Wu, Xiaoli Xu,~\IEEEmembership{Member, IEEE},
and 
Yong Zeng,~\IEEEmembership{Senior Member, IEEE}
  } 
  \thanks{          Z. Dai, D. Wu,  X. Xu, and Y. Zeng  are with the National Mobile Communications Research Laboratory, 
  Southeast University, Nanjing 210096, China. Y. Zeng is also with the 
  Purple Mountain Laboratories, Nanjing 211111, 
  China (e-mail: \{zhuoyin\_dai, studywudi,  xiaolixu, yong\_zeng, \}@seu.edu.cn). 
  (\emph{Corresponding author: Yong Zeng.})
} 
  
  }

\markboth{Journal of \LaTeX\ Class Files,~Vol.~14, No.~8, November~2024}%
{Shell \MakeLowercase{\textit{et al.}}: A Sample Article Using IEEEtran.cls for IEEE Journals}


 \maketitle


\begin{abstract}
  Channel knowledge map (CKM) is 
  a promising paradigm shift towards environment-aware communication and sensing by
  providing location-specific prior channel knowledge before real-time communication.
  Although CKM is particularly appealing for  dense networks such as cell-free networks,
  it remains  a challenge to efficiently generate CKMs in dense networks.
  For a dense network with CKMs of existing access points (APs), 
  it will be useful to  efficiently generate CKMs of  potentially new APs with only AP location information.
  The generation of inferred CKMs across APs can help dense networks achieve 
  convenient initial CKM generation, environment-aware AP deployment, and cost-effective 
  CKM updates.
  Considering that  different   APs in the same region
  share the same  physical environment, there exists a 
  natural   correlation between the channel knowledge of different APs. 
  Therefore, by mining the implicit correlation between location-specific 
  channel knowledge, cross-AP CKM inference can be 
  realized using data from other APs.
  This paper proposes a cross-AP  inference method to generate CKMs of  potentially new APs
  with deep learning.
  The location of the target AP is fed into the UNet model in combination with 
  the channel knowledge of other existing APs, and supervised learning is 
  performed based on the channel knowledge of the target AP.
  Based on the trained UNet and the channel knowledge of the existing APs,
  the CKM inference of the potentially new AP can be generated across APs.
  The generation results of the inferred CKM
  validate the feasibility and effectiveness of cross-AP CKM inference with other APs' channel knowledge.

\end{abstract}
   

\section{Introduction}


As a promising paradigm shift from conventional environment-unaware 
to  environment-aware communication and sensing, channel knowledge map (CKM) has been recently 
proposed to address the challenge of channel knowledge acquisition with the prior local environment \cite{CKM,Zeng2023ATO}.
As a location-specific channel knowledge database, CKM can effectively reflect the   channel knowledge 
related to the   node location and dependent on the local environment, such as channel gain, time of arrival (ToA),
angle of arrival (AoA), angle of departure (DoA), etc \cite{wu2022environment}.
Different from the physical environment map \cite{seidel1994site}, CKM focuses   on the intrinsic characteristics of wireless channels, 
which effectively avoids the complicated computation from  physical environment 
to channel knowledge. Therefore, the prior local environment 
embedded in CKMs can greatly facilitate the performance optimization and 
resource allocation of future wireless communications.

Efficient construction of CKM is the key to realizing CKM-based environment-aware communication and sensing.
Essentially, the construction of CKM is the process of combining limited   
data with prediction methods 
such as interpolation  and inference and  obtaining the location-specific 
channel knowledge in the region \cite{Zeng2023ATO}.  
CKM construction based on existing channel knowledge can be categorized into same-AP construction and cross-AP construction.
For the same AP, the CKM can be completed or the CKM  resolution can be improved 
by using the physical environment map \cite{levie2021radiomao} or the 
channel knowledge of the nearest neighbor nodes \cite{xu2023much}. 
For the cross-AP construction, the mutual information proves the 
existence of channel state information (CSI)  dependence across APs, and CSI features such as received power 
at a specific location are inferred from the source CSI \cite{chen2019learning}.
Learning-based channel mapping is also used for cross-antenna channel prediction 
at specific candidate locations \cite{Alrabeiah2019deep}.

Different from the CSI inference for a specific location across APs/antennas above, 
this letter focuses on the fundamental problem that constructing the complete  CKMs 
of potentially new APs efficiently based on the 
CKMs of existing APs. Specifically, consider a dense network, such as a cell-free network, in which the existing APs 
are equipped with CKMs for the region.
A potentially new AP is introduced into the network with 
only its location information  known. 
A cross-AP CKM generation system needs to be designed whose inputs are the 
location information of the potentially new AP with the CKMs of other APs 
in the network, while the output is the complete CKM inference of the new AP.
With the densification of network nodes, the generation 
of the cross-AP inferred CKMs of potentially new APs  effectively expands the system potential.
The overhead of constructing CKMs for all APs can be reduced 
during the initial CKM construction phase through a combination of measurement and
cross-AP inference strategies.
For newly introduced potential APs, traversing to generate CKMs in different locations 
can guide the environment-aware placement.
CKM inference across APs also realizes cost-effective  CKM updates during subsequent system maintenance.

\begin{figure}[!t]
  \centering
    {\includegraphics[width=0.75\columnwidth]{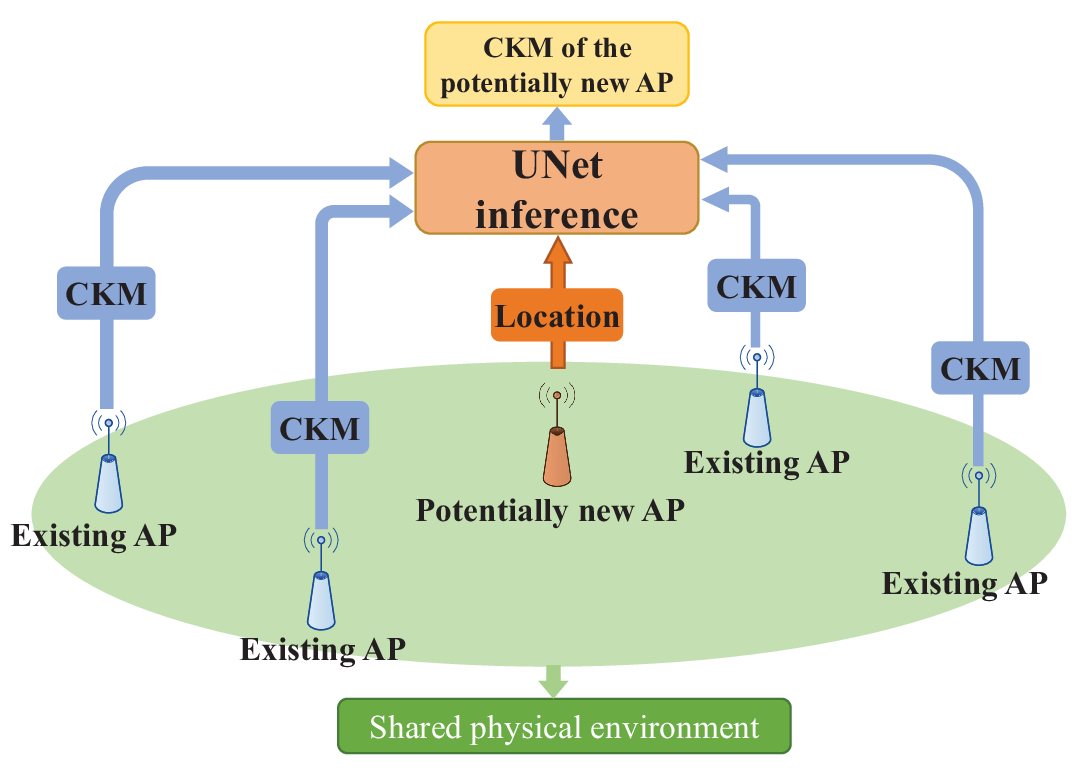}}%
  \caption{Model of the  Cross-AP CKM Inference.}
  \label{fig:mapping}
\end{figure}

The cross-AP CKM inference is built on the location diversity of distributed APs
and the same physical environment they share.
Specifically, the wireless environment is an outward manifestation of the physical environment, 
which is also embedded in the spatial variations of the wireless environment \cite{Zeng2023ATO}.
Therefore, there is a natural correlation and dependence between the CKMs of distributed APs 
at different locations.
Although this implicit relationship is difficult to characterize concretely, 
it can be   applied to cross-AP CKM inference with learning-based approaches.
In this paper, neural network is used to learn the implicit correlation between CKMs 
of different APs related to the wireless environment, and ultimately to 
realize cross-AP  CKM inference. 
As shown in Fig. \ref{fig:mapping}, this learning-based cross-AP CKM inference avoids the 
complex computation from physical environment to channel knowledge.
The UNet model \cite{ronneberger2015u} is first trained with CKM datasets 
from different physical environments. After training, the locations of
the potentially new APs are fed into the model along with the CKMs of other existing
APs  to output the CKMs of the potentially new APs.
This cross-AP CKM inference learns the correlation between CKMs and 
is capable of generating complete CKMs of potentially new APs with high accuracy.
The generated CKM results are compared with other benchmarks 
to validate the feasibility and effectiveness of CKM inference across APs.

\section{  Problem Formulation}
As shown in  Fig. \ref{fig:mapping}, consider a cell-free network with 
 $N$  APs.
The coordinate of the $n$th AP is denoted as $\mathbf{c}_n \in \mathbb{R}^{2\times 1}$. 
The problem to be solved is:  how to construct a CKM for a potentially new AP at location  $\mathbf{c}_0$ based on the CKMs of existing APs in the cell-free network?
Specifically, the problem is first analyzed with a typical kind of CKM, channel gain map (CGM), as an example.
The entire network area is divided into $W \times W$ grids, where each grid records a channel gain value.
Therefore, for the $n$th AP, its CKM, which mainly stores the location of the AP itself and the channel gain 
corresponding to each grid, can be expressed as
\begin{equation}\label{eq:CKMstore  }
  \mathcal{M}_n=\{\mathbf{c}_n;\mathbf{G}_n\},
\end{equation}
where   
$\mathbf{G}_n\in \mathbb{R}^{W\times W}$ denotes the  channel gain   of the $n$th AP.

Consider a potentially new AP 0 for which only its location information $\mathbf{c}_0$ is known. 
The focus of this paper is on how to reduce or even avoid actual measurements 
to generate the CKM  for  the AP 0. 
Based on the correlation on the physical environment, CKMs of other APs within the network area 
are naturally a source of information that can be mined.
Therefore, the CKM construction problem for AP 0 can be formulated as a cross-AP  inference problem 
from  CKMs  of other existing distributed APs to the CKM of AP 0 as 
\begin{equation}\label{eq:mapping}
  f: \big\{ \mathcal{M}_n \big\}_{n=1}^{N};\mathbf{c}_0\rightarrow \mathbf{G}_0 .
\end{equation}

\begin{figure}[!t]
  \centering
    {\includegraphics[width=0.85\columnwidth]{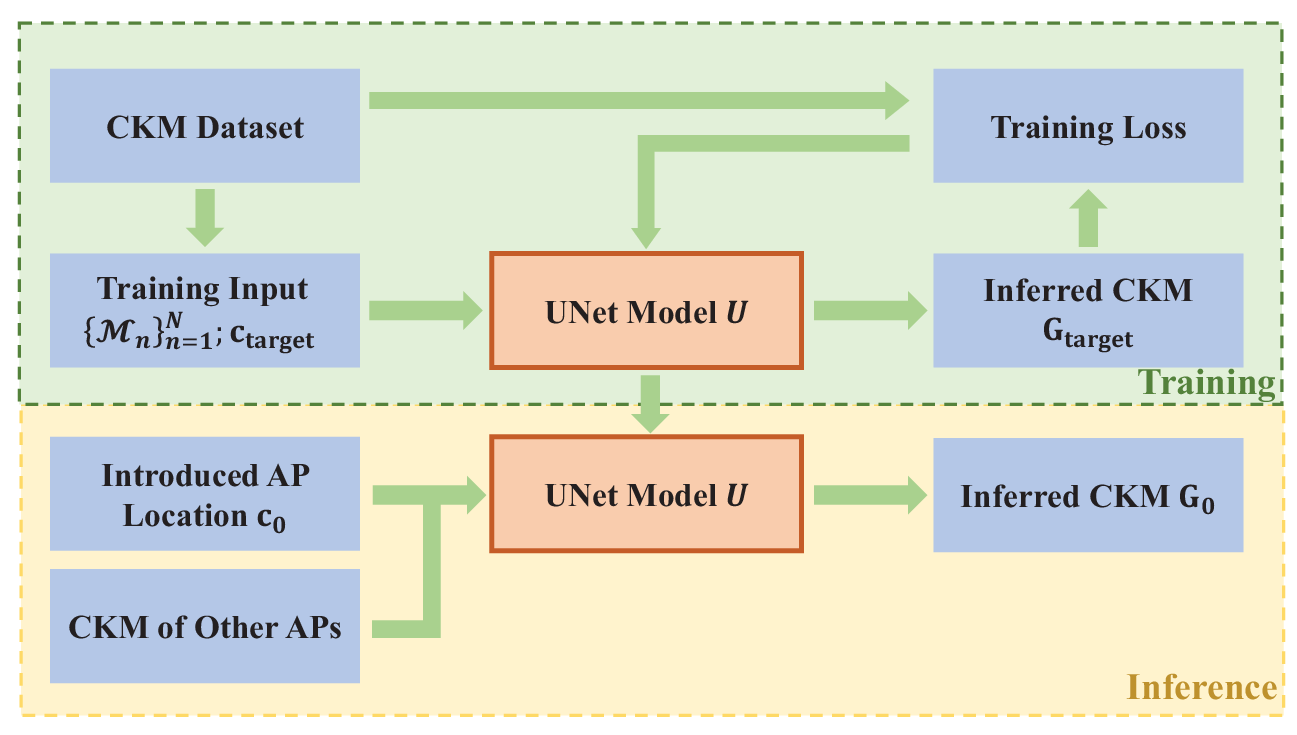}}%
  \caption{Illustrating of Different Phases of  the Cross-AP CKM Inference.}
  \label{fig:phase}
\end{figure}
The training and inference phases of cross-AP CKM inference are illustrated 
in Fig \ref{fig:phase}.
In the training phase, UNet learns the propagation characteristics in the 
wireless environment by performing supervised learning with the help 
of the channel knowledge database and optimizing the UNet parameters.
In the inference phase, the coordinate $\mathbf{c}_0$ of the introduced AP 0 
and the CKMs of other APs in the area are combined and input into the 
trained UNet model, and the final inferred CKM $\mathbf{G}_0 $ is output.

The proposed  CKM inference method across APs  is based on the fact that  
the physical environment 
shared by the  APs determines the wireless channel conditions. We believe that the 
cross-AP CKM inference has great potential in cell-free networks.
In the initial phase of the  network deployment, the cross-AP  CKM inference 
can effectively reduce the overhead of CKM construction.
Specifically, CKMs are mapped for only part of the APs through actual measurements 
or ray tracing, while CKMs for the remaining APs are directly inferred 
to effectively reduce the overall construction overhead.
Meanwhile, in a cell-free network already equipped with CKMs, cross-AP CKM inference 
can effectively guide the environment-aware deployment of potentially new APs.
By inferring the corresponding CKMs, the coverage of the
potentially new AP in different locations can be effectively modeled, 
so that the locations where the wireless environment best meets the requirements 
can be selected for AP deployment.

\section{Model Training}

In this section, the construction and training of the UNet  for 
cross-AP CKM inference is presented.
Specifically, CKMs from distributed APs within the cell-free network are first 
transformed and combined   to generate input data for UNet training.
Then, the design of the UNet architecture for cross-AP CKM inference 
 is presented.

\subsection{Input Data Generation}

The physical and wireless environments interact with each other.
As an example, abrupt variations in channel strength in space often imply abrupt 
variations in the wireless environment, which can be further inferred from variations 
in the physical environment due to obstacles, buildings, etc.
Therefore, effective cognition of the wireless environment can be obtained by 
synthesizing the CKMs of distributed APs in cell-free networks.
Although this cognition of the wireless environment is difficult to express directly 
in a concrete mathematical form, it can be learned through neural networks.
Specifically, the AP location stored in each CKM is first converted into a 
$W\times W$ AP location map through one-hot encoding, i.e., for any $n$th AP, there is
\begin{equation}
  \mathbf{M}_{\rm{AP},n}(\mathbf{c})=\left \{ 
    \begin{aligned}
      1, \quad \mathbf{c}=\mathbf{c}_n \\
      0, \quad \mathbf{c}\neq \mathbf{c}_n,
    \end{aligned}\right 
    .
\end{equation}
where $\mathbf{c}$ is any poss coordinate in the CKM.

Combining AP location maps with their corresponding CGMs can be a good way to help UNet learn wireless environment characteristics.
In general,   the grid of the AP  
tends to have the maximum value of channel gain in $\mathbf{G}_n$.
To highlight the AP location feature and strengthen the influence of AP location on CKM, 
the AP location map $\mathbf{M}_{\rm{AP},n}$ is  considered to be weighted 
with the corresponding channel gain matrix $\mathbf{G}_n$ as the feature map of the $n$th AP
\begin{equation}
  \mathbf{M}_{n}=(1-\omega)\mathbf{G}_n +\omega \mathbf{M}_{\mathrm{AP},n},
\end{equation}
where increasing the weight $\omega$ strengthens the corresponding 
AP location feature but weakens the details of the CKM.

Further, combining $\mathbf{M}_{n}$ of all the $N$ distributed APs in the cell-free network 
results in a feature map of $N$-dimensional channels, where  the feature map 
of each AP corresponds to one dimension.
Name the AP of which the CKM needs to be generated as the target AP.
The next part is to combine the target AP location map with the $N$-dimensional 
feature maps of the other APs.
Notice that the AP location information stored in map $\mathbf{M}_{\rm{AP},{\rm target}}$ is sparse.
To make the convolutional kernel fully capture the key AP location features and 
avoid feature omissions during the learning process,  
pre-convolution is first needed to reinforce the features of the 
$\mathbf{M}_{\rm{AP},{\rm target}}$.
By convolving with the  kernel $\mathbf{v}=\mathbf{1}\in \mathbb{R}^{3\times 3}$, 
 all elements in the surrounding $3\times 3$ grids of the 
AP location $\mathbf{c}_{\rm{target}}$ are set to 1, effectively expanding the coverage of 
location information.
The pre-convolution kernel can 
improve the extraction of AP location features, and make its performance more 
stable in the face of sparse location data. The target AP location map after pre-convolution 
is
\begin{equation}
  \mathbf{M}^{*}_{\rm{AP},{\rm target}}=\mathbf{M}_{\rm{AP},target}\ast \mathbf{v}.
\end{equation} 

\begin{figure}[!t]
  \centering
    {\includegraphics[width=0.75\columnwidth]{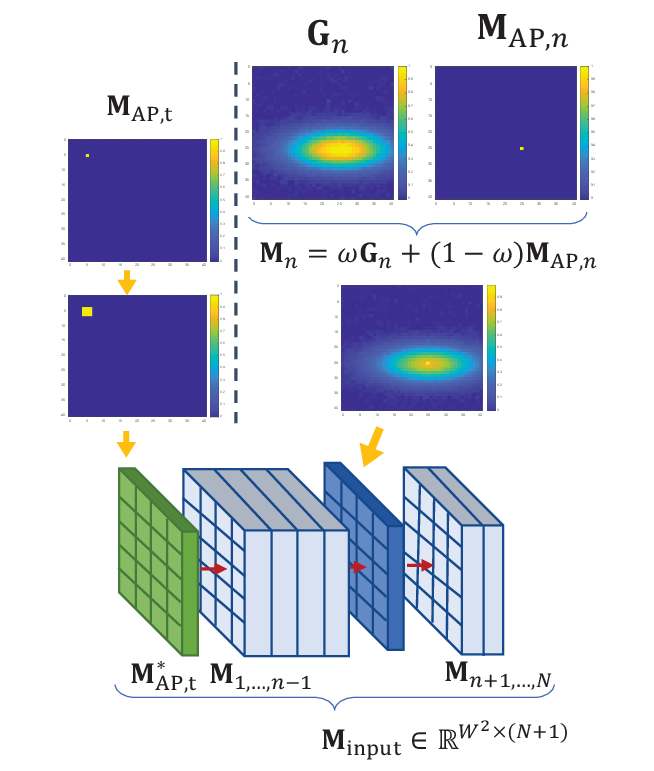}}%
  \caption{ Input Structure Design.}
  \label{fig:inputdata}
\end{figure}

Finally, the target AP location map $\mathbf{M}^{*}_{\mathrm{AP},{\rm target}}$ 
is overlaid to generate the input data $ \mathbf{M}_{\mathrm{input}}\in \mathbb{R}^{W^2 \times (n+1)}$
of the UNet network.
The generation process and structure of  $ \mathbf{M}_{\mathrm{input}}$ is shown in Fig. \ref{fig:inputdata}.

\subsection{UNet  Design and Training}
The structure of the UNet network for cross-AP inference is designed 
based on the CGMs and their corresponding AP location maps in 
RadioMapSeer \cite{DatasetPaper}. For other datasets like CKMImageNet \cite{wudi2024ckmimagenet}, parameters such as 
the number of UNet input channels need to be adjusted according
to the characteristics of the dataset. 

\begin{figure*}
  \centering
  \includegraphics[width=0.9\textwidth]{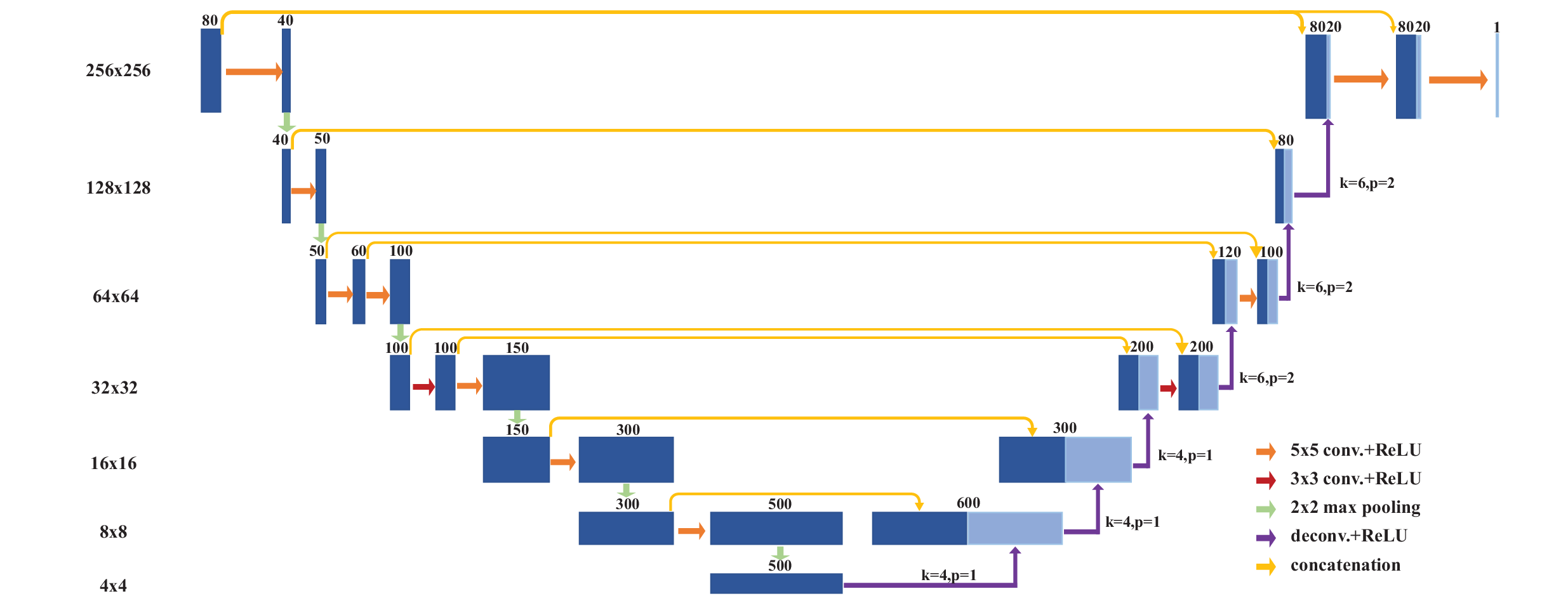}
  \caption{Structure of the UNet for Cross-AP CKM Inference. \hfill}
  \label{fig:UNet}
  \end{figure*}

The details of the UNet structure are shown in Fig. \ref{fig:UNet}.
Note that for each physical environment map in RadioMapSeer, there are 80 
AP location maps and corresponding simulated CGMs. Therefore, the input data of the 
UNet  for cross-AP inference is a high-dimensional matrix with 80 channels.
According to the process of input data generation, the first  channel dimension 
consists of the target AP location map $\mathbf{M}^{*}_{\rm{AP},target}$ after mask operation.
All the other channels are feature maps composed of the weighted sums of CGMs and AP 
location maps of other APs in the same physical environment map.
As a supervised learning, the output data in the UNet architecture is the inferred
CGM $\mathbf{G}^{\rm{infer}}_{\rm{target}}$ of the target AP with one channel.

The network architecture has the symmetry of the classical UNet structure and can be divided into a downsampling part and an upsampling part.
To enhance the learning of wireless environment features at the edges of the building, 
the $5\times5 $ convolution kernel is used extensively, which helps to capture the 
local spatial information.
For cross-AP CKM inference, the high-channel input data can lead to parameter redundancy. 
Dimensionality-reduction convolutions effectively reduce the parameter redundancy, 
significantly lowering computational overhead and improving model generalization.
Further, in the process of down-sampling and up-sampling, the network extracts 
and recovers increasing high-level features layer by layer, while retaining 
low-level features with hop concatenation.
Meanwhile, additional convolutions are added at the resolutions of $64\times64 $ and $32\times32$, 
which helps to enrich and strengthen the local feature representation 
and  the CKM property characterization.

During the training process,  each AP  randomly takes turns to play the role 
of target AP, while the remaining  APs 
act as the other existing APs with CKMs.
The input data $ \mathbf{M}_{\mathrm{input}}$ is generated and fed into the UNet
and outputs the corresponding inferred CKM $\mathbf{G}^{\rm{infer}}_{\rm{target}}$.
The corresponding   CGM $\mathbf{G}_{\rm{target}}$ of the target AP is the ground-truth. 
The  mean square error (MSE) between   $\mathbf{G}_{\rm{target}}$
and    $\mathbf{G}^{\rm{infer}}_{\rm{target}}$  is used as the loss function as
\begin{equation}
  e=\frac{1}{W^2} \sum_{i=1}^{W^2}( \mathbf{G}_{\rm{target}}(i)-\mathbf{G}^{\rm{infer}}_{\rm{target}}(i)  )^2.
\end{equation}
By updating the parameters according to the loss, 
the trained UNet model can be finally obtained.
The detailed process of the training phase is  shown in  Algorithm \ref{algorithm:1}.

\begin{algorithm}
  \caption{Training phase of the cross-AP inference}\label{algorithm:1}
  \KwData{the training CKM set $\{ \mathcal{M}_n \big\}_{n=1}^{N}$}
  \KwResult{the optimal parameters $\{ \bm{\theta^{\star}} \}$}
  Get the AP location map $\mathbf{M}_{\rm{AP},n}$ for any $n$ in dataset\;
  \For{\rm{each epoch}}{
    \For{ \rm{each target AP in the training dataset }}
    {
      Get $\mathbf{M}^{*}_{\rm{AP},{\rm target}}$ through  pre-convolution\;
      Get the  feature map $\mathbf{M}_{n}$ of all the other  APs\;
      Generate the input data $ \mathbf{M}_{\mathrm{input}}$\;
      Input $ \mathbf{M}_{\mathrm{input}}$ into the UNet and get $\mathbf{G}^{\rm{infer}}_{\rm{target}}$\;
     Calculate the loss and gradients with  $\mathbf{G}_{\rm{target}}$\;
      Update the UNet parameters $\{ \bm{\theta} \}$
    }

  Save the parameters $\{ \bm{\theta^{\star}} \}$ with the best loss\;

  }
  \Return{\rm{the optimal parameters} $\{ \bm{\theta^{\star}} \}$}
  \end{algorithm}


\section{Inference Results}
In this section, the UNet network is trained based on  RadioMapSeer Dataset.
The trained UNet network will perform cross-AP CKM inference for validation.
The inference results for CKM will be compared with the benchmark schemes 
to demonstrate the feasibility of generating  inferred CKM  across APs in 
cell-free networks without physical environment.

\subsection{Training Settings}
To ensure the independence between  the training dataset and the validation dataset, 
500 of  the 700 different physical environment maps  in RadioMapSeer are 
 arbitrarily selected for model training.
The training dataset consists of 80 AP location maps and their CGMs corresponding to each 
physical environment map and does not contain the physical environment maps themselves.
During the training process, each AP of each physical map acts as the target AP in turn 
and uses the corresponding target CGM as the ground-truth, constituting a 
total of 40000 sets of input-output training data.
The training is performed based on Adam, 
 where the 
initial learning rate is $10^{-3}$. 
The UNet  training is carried out for 15 epochs, where the batch size is 15.
To mitigate overfitting, the model that minimized the MSE loss in the validation set 
over the 15 epochs is saved.

\subsection{Training Results}

After training, CKM inference across APs was performed on 
the validation dataset with the UNet model.  
The 100 physical environment maps in RadioMapSeer Dataset that are disjoint 
from the training dataset are selected as the validation dataset.
Similarly, the validation dataset is composed of 80 AP location maps with CGMs 
for each physical environment map. 
Two basic CKM construction methods are chosen as benchmarks.
\begin{itemize}
  \item  Weighted cross-AP CKM inference scheme. The inferred CKM  
  is represented as a weighted sum of the CKMs of the other $N$ APs.
  
  \begin{equation}
    \mathbf{G}^{\rm{infer}}_0=\sum_{n=1}^{N} w_{n}\mathbf{G}_n
    =\sum_{n=1}^{N} \frac{e^{-\beta d_{t,n}}}{\sum_{i=1}^{N} e^{-\beta d_{t,i}} }\mathbf{G}_n,
  \end{equation}
  where $w_{n}$ denotes the weight  based on the Euclidean distance, $\beta=0.1$ denotes the weight parameter. The distance between the target AP and 
  the  AP $i$ is $d_{t,i}$.
  \item Path loss model in urban microcell   proposed by 3GPP TR 38.901 \cite{3gpp.38.901}.
\end{itemize}

\begin{figure*}[t]
    \centering
    \subfigure[CKM Ground-truth(simple)]{\label{fig:subfig:a}
    \includegraphics[width=0.30\columnwidth]{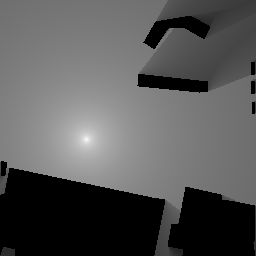}}
    \subfigure[Proposed CKM Inference(simple)]{\label{fig:subfig:b}
    \includegraphics[width=0.30\columnwidth]{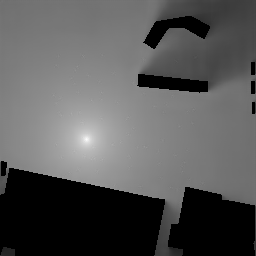}}
     {\color{blue}\rule{2pt}{2.5cm}}
    \subfigure[CKM Ground-truth(medium)]{\label{fig:subfig:c}
    \includegraphics[width=0.30\columnwidth]{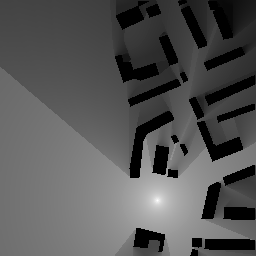}}
    \subfigure[Proposed CKM Inference(medium)]{\label{fig:subfig:d}
    \includegraphics[width=0.30\columnwidth]{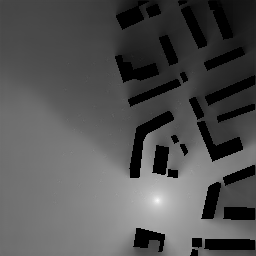}}
    {\color{blue}\rule{2pt}{2.5cm}}
    \subfigure[CKM Ground-truth(complex)]{\label{fig:subfig:e}
    \includegraphics[width=0.30\columnwidth]{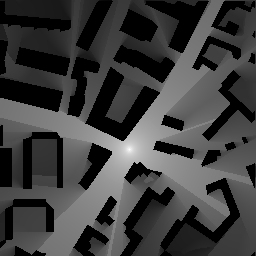}}
    \subfigure[Proposed CKM Inference(complex)]{\label{fig:subfig:f}
    \includegraphics[width=0.30\columnwidth]{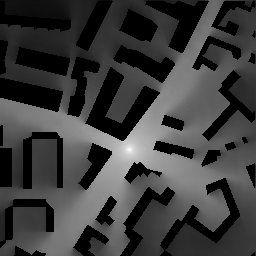}}
    \vfill
    \subfigure[Weighted  CKM Inference(simple)]{\label{fig:subfig:g}
    \includegraphics[width=0.30\columnwidth]{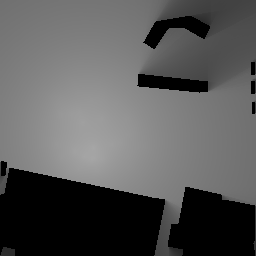}}
    \subfigure[Model-based CKM    Inference(simple)]{\label{fig:subfig:h}
    \includegraphics[width=0.30\columnwidth]{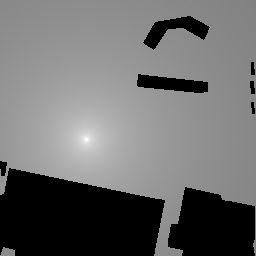}}
    {\color{blue}\rule{2pt}{2.5cm}}
    \subfigure[Weighted  CKM Inference(medium)]{\label{fig:subfig:i}
    \includegraphics[width=0.30\columnwidth]{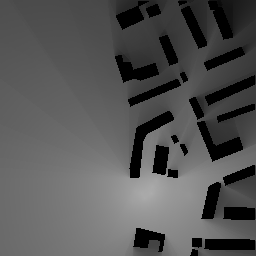}}
    \subfigure[Model-based  CKM Inference(medium)]{\label{fig:subfig:j}
    \includegraphics[width=0.30\columnwidth]{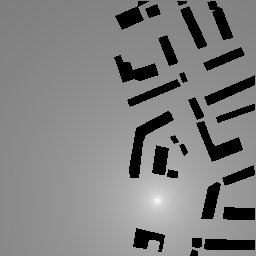}}
    {\color{blue}\rule{2pt}{2.5cm}}
    \subfigure[Weighted  CKM Inference(complex)]{\label{fig:subfig:k}
    \includegraphics[width=0.30\columnwidth]{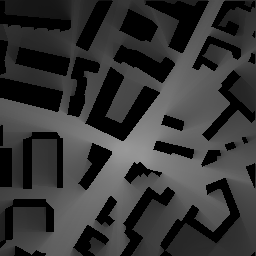}}
    \subfigure[Model-based  CKM Inference(complex)]{\label{fig:subfig:l}
    \includegraphics[width=0.30\columnwidth]{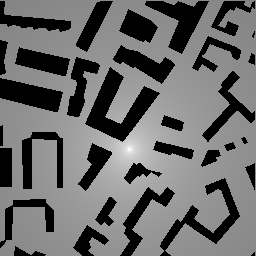}}
    \caption{Comparison of the  Cross-AP CKM Inference and the  Benchmarks.}
    \label{fig:subfig}
    \end{figure*}

The MSEs and root MSEs (RMSEs) of different cross-AP CKM inference schemes are shown in 
Table I. Note that the  range of the channel gain from the noise floor to
the maximum in RadioMapSeer Dataset is 100dB, the unit of the RMSE is dB.
As shown in Table I, the accuracy of the learning-based cross-AP CKM inference is about 2.38 dB.
This mean error is on the same level as that of RadioUNet(2.03dB), 
but without the need for 
the physical environment map as training data.
Compared with the benchmark schemes, the accuracy of the proposed cross-AP CKM 
inference is   3dB higher than the distance-based weighted inference, 
and 33dB higher than the model-based inference.
The comparisons of the cross-AP CKM inference and the benchmark schemes
are presented in Fig. \ref{fig:subfig}. 
The  inference is performed in different physical 
environments ranging from simple to complex.
The distance-based weighted inference  blurs key features of the 
CKM such as the AP location.
In contrast, a comparison with the CKM ground-truth reveals that 
the proposed cross-AP CKM 
inference  well preserves the target AP location and learns the wireless environment 
features and wireless propagation characteristics. 
Even unaware of the physical environment map, the CKM inference still exhibits 
the attenuation and mutation characteristics  under the occlusion of buildings.

\begin{table}[H]
  \caption{Cross-AP CKM  Inference MSE}
  \label{table1p}
  \centering
  \begin{tabular}{cccc}
  \toprule
  Scheme &MSE(dB$^2$) & RMSE(dB) \\\midrule
   Proposed cross-AP CKM inference  &5.66&2.38\\
  RadioUNet CKM generation \cite{levie2021radiomao}   &4.12 & 2.03 \\
  Benchmark 1: weighted CKM inference  & 28.04&5.30\\
  Benchmark 2: model-based CKM inference  & 1275.58&35.72\\
  \bottomrule
  \end{tabular}
  \end{table}

 \section{Conclusion} 
 
 This paper proposes a  cross-AP CKM inference   in cell-free
 networks. By taking advantage of the correlation of the wireless environment
 and  the shared physical environment among APs, the trained UNet  
 utilizes   other existing APs' CKMs to generate CKMs of potentially 
 new APs without the need for physical environment maps or any onsite measurement.
 The comparisons with the CKM inference by benchmark schemes validate the  feasibility and effectiveness 
  of cross-AP CKM inference, which is significant for the construction and 
 updating of   CKMs as well as the environment-aware deployment of potentially new APs in dense networks.

 \begin{appendices}

\end{appendices}


\bibliographystyle{IEEEtran}

\bibliography{IEEEabrv,ref}


\end{document}